\documentclass[12pt]{article}

\usepackage{graphicx}

\begin{document}

\begin{center}
{\bf Nonlinear electrodynamics and black holes} \\
\vspace{5mm} S. I. Kruglov
\footnote{E-mail: serguei.krouglov@utoronto.ca}

\vspace{3mm}
\textit{Department of Chemical and Physical Sciences, University of Toronto,\\
3359 Mississauga Road North, Mississauga, Ontario L5L 1C6, Canada} \\
\vspace{5mm}
\end{center}

\begin{abstract}
We investigate a new model of nonlinear electromagnetic field coupled with the gravitation field. The black hole solution is obtained possessing the asymptotic Reissner-Nordstr\"om solution.
The electric field has the finite value at the origin and there are not singularities.
%We show that cosmological constant appears naturally depending on the Planck mass $M_{Pl}$ and the dimensional
%parameter $\beta$ of the nonlinear electromagnetic field, $\Lambda=1/(2M_{Pl}^2\beta)$.
\end{abstract}

\section{Introduction}

Cosmological models with the use of nonlinear electrodynamics (NLE) can solve some problems
such as the initial Big Bang singularity and early time inflation. This is because some models of NLE
are without divergences and give finite self-energy of charged particles and can remove singularity at the classical level. NLE models are effective models that take into account quantum corrections. Thus, Maxwell Lagrangian plus
Heisenberg-Euler Lagrangian \cite{Heisenberg} yield NLE that takes into account one-loop quantum corrections to QED. The Born-Infeld theory \cite{Born} is NLE which smoothes divergences and can be explored for strong electromagnetic fields as it gives the restriction on the possible maximum electric field. Cosmological models use the classical Einstein equation and, therefore, the classical NLE fields may be included in the gravity theory. It is of interest to consider the electromagnetic radiation and gravitational field in the early epoch of the universe because the fields at that epoch are very strong, and the nonlinear electromagnetic effects are very important. Thus, NLE models are interesting in general relativity (GR), and in cosmological theories, as a simple classical models that take into consideration vacuum polarization processes and can influence on the evolution of the early universe near the Planck era. In addition, nonlinear electromagnetic fields can mimic the dark energy in some inflationary models.
High energies in early universe can stimulate the appearance of nonlinear electromagnetic effects.
In early epochs of the universe the magnetic fields can be greater than $10^{15}$ G. At these values of the strength of the electromagnetic field the self-interaction of photons is important and classical electrodynamics should be modified \cite{Jackson}. This motivates to use spacetimes with nonlinear electromagnetic fields. One may consider the Maxwell fields as an approximations of NLE theory for weak fields. Generalizations of Maxwell theory to NLE were introduced to eliminate infinite quantities. In late epochs the nonlinear electromagnetic fields may be considered as an effective and such phenomenological approach \cite{Medeiros} mimics a material media with electric
permittivity and magnetic permeability depending on the fields \cite{Plebanski}. So, nonlinear electrodynamics has been explored to produce inflation in the early universe \cite{Garcia}, \cite{Camara}. Some models were considered to use nonlinear electromagnetic fields to have accelerated expansion \cite{Elizalde}-\cite{Novello1}.
The effect of coupling nonlinear electrodynamics to gravity can produce negative pressures that accelerate the expansion \cite{Novello}-\cite{Vollick}. The popular point of view is that the cosmological constant drives the present cosmic acceleration.
There is a parallel between the trace anomaly of NLE theory and cosmological constant \cite{Labun}.
The Einstein-Born-Infeld equations were studied since they take into account nonlinear effects in strong electromagnetic and gravitational fields \cite{Hendi}. In this paper we consider the NLE model, which is self-consistent and satisfies all natural requirements, coupled to gravitational field. That NLE equations depend on a dimensional constant $\beta$ giving a maximum electric field strength \cite{Kruglov}. Our goal is to investigate Einstein-NLE model and its influence on the evolution of universe. Such model may also prevent the initial big bang singularity. The trace of the energy-momentum tensor $T$ can make contribution to the cosmological constant \cite{Schutzhold}. Thus, in curved spacetimes nonperturbative effects of self-interacting quantum fields may give a contribution to the cosmological constant. GR can be considered as an effective field theory at low energy of quantum gravity theory so that, the Einstein-Hilbert classical action of GR should have the energy-momentum tensor including the trace anomaly \cite{Mottola}. The breaking of scale invariance in NLE can have the result that the equation of state allows for negative pressure. The Reissner-Nordstr\"{o}m (RN) solution describes the black hole with the charge $Q$ and mass $M$ and it can be the final state of charged stars. We investigate in details its nonlinear generalization exploring new NLE model. In this paper we study the Einstein-NLE solution that is the nonlinear electromagnetic generalization of the RN black hole. We consider the static and spherically symmetric charged black holes with NLE as a source. The solutions obtained generalize the RN geometry. The black hole geometry in the presence of a nonlinear electromagnetic field reduces in the weak-field limit to Einstein-Maxwell geometry. The static and spherically symmetric spacetime of black hole with the Euler-Heisenberg effective Lagrangian of QED as a source was learned in \cite{Hiroki}.

We use the units in which $c=\hbar=1$.
The nonlinear electromagnetic field introduced in \cite{Kruglov} is given by the Lagrangian density
\begin{equation}
{\cal L}_{em} = -\frac{{\cal F}}{2\beta{\cal F}+1},
 \label{1}
\end{equation}
where $\beta$ is dimensional parameter with $\beta{\cal F}$ being dimensionless, ${\cal F}=(1/4)F_{\mu\nu}F^{\mu\nu}$, and $F_{\mu\nu}$ is the field strength. The parameter $\beta$, with the dimension of (length)$^4$, is connected with the upper bound on the possible electric field strength. The energy-momentum tensor is as follows \cite{Kruglov}:
\begin{equation}
T_{\mu\nu}=-\frac{1}{(2\beta{\cal F}+1)^2}\left[F_{\mu}^{~\alpha}F_{\nu\alpha}-g_{\mu\nu}{\cal F}(2\beta{\cal F}+1)\right],
 \label{2}
\end{equation}
and has non-vanishing trace. In the model based on Eq. (1) the electric field of a point-like charge
does not have singularity at the origin, and the maximum possible electric field is $E_{max}=1/\sqrt{\beta}$. There is also the finiteness of the electromagnetic energy of a point-like particle and the mass of the particle can be treated as a pure electromagnetic energy.

\section{Nonlinear electromagnetic field and black holes}

Let us consider the action of GR coupled with the nonlinear electromagnetic field (1)
\begin{equation}
S=\int d^4x\sqrt{-g}\left[\frac{1}{2\kappa^2}R+ {\cal L}_{em}\right],
\label{3}
\end{equation}
where $\kappa^{-1}=M_{Pl}$, $M_{Pl}$ is the reduced Planck mass, $R$ is the Ricci scalar.
Einstein's and electromagnetic equations, obtained from Eq. (3), are given by
\begin{equation}
R_{\mu\nu}-\frac{1}{2}g_{\mu\nu}R=\kappa^2T_{\mu\nu},
\label{4}
\end{equation}
\begin{equation}
\partial_\mu\left(\frac{\sqrt{-g}F^{\mu\nu}}{(2\beta{\cal F}+1)^2}\right)=0.
\label{5}
\end{equation}
We are going to find the static charged black hole solutions to Eqs. (4),(5). The spherically symmetric line element in $(3+1)$-dimensional spacetime is
\begin{equation}
ds^2=-f(r)dt^2+\frac{1}{f(r)}dr^2+r^2(d\vartheta^2+\sin^2\vartheta d\phi^2).
\label{6}
\end{equation}
We suppose that the vector-potential has non-vanishing component $A_0=h(r)$ so that ${\cal F}=-(h'(r))^2/2$, where the prime means the derivative with respect to the argument. Then Eq. (5) becomes
\begin{equation}
\partial_r\left(\frac{r^2 h'(r)}{[1-\beta (h'(r))^2]^2}\right)=0.
\label{7}
\end{equation}
From Eq. (7), after integration, we obtain
\begin{equation}
r^2h'(r)=C[1-\beta (h'(r))^2]^2,
\label{8}
\end{equation}
where $C$ is the dimensionless integration constant.
Introducing new dimensionless variables
\begin{equation}
y=\frac{r\sqrt{h'(r)}}{\sqrt{C}},~~~~x=\frac{r}{\sqrt{C}\beta^{1/4}},
\label{9}
\end{equation}
we find from Eq. (8) the algebraic equation
\begin{equation}
y^4+(y-1)x^4=0.
\label{10}
\end{equation}
With the help of Cardano's formulas one can write down the analytic solution to Eq. (10). The plot of the function $y(x)$ is presented in Fig. 1.
\begin{figure}[h]
\includegraphics[height=4.0in,width=4.0in]{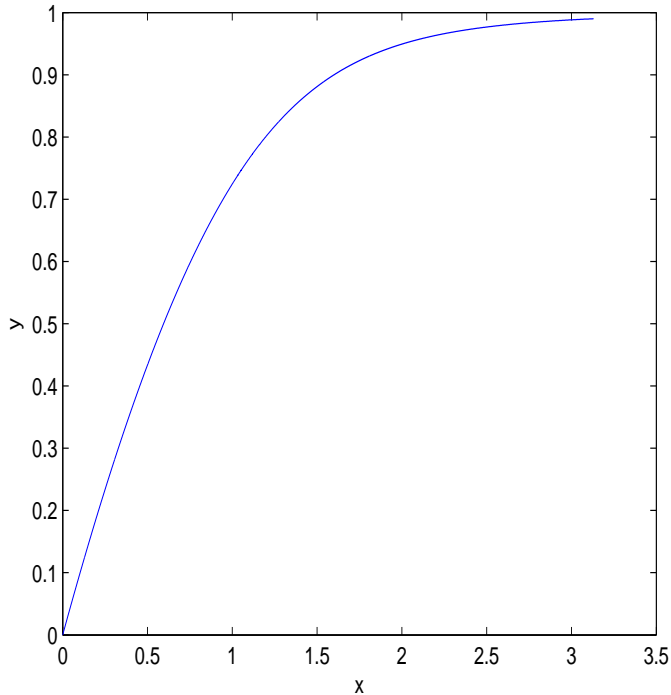}
\caption{\label{fig.1}The function  $y$  versus $x$.}
\end{figure}
From Eq. (10) we obtain the asymptotic values $r\rightarrow 0$
($x\rightarrow 0$), $y\rightarrow 0$, and when $r\rightarrow \infty$
($x\rightarrow \infty$), $y\rightarrow 1$. Now we can find from Eq. (9) the integral
\begin{equation}
h(r)=C\int \frac{y^2}{r^2}dr=\frac{\sqrt{C}}{4\beta^{1/4}}\int \frac{(4-3y)dy}{(1-y)^{3/4}}=\frac{\sqrt{C}}{5\beta^{1/4}}(3y-8)(1-y)^{1/4},
\label{11}
\end{equation}
where $y$ is the solution to Eq. (10) and $x$, as a function of $r$, is given by Eq. (9).
From Eq. (11) we obtain
\begin{equation}
h'(r)=\frac{\sqrt{1-y}}{\sqrt{\beta}}.
\label{12}
\end{equation}
As a result, according to Eq. (11), (12) at $r\rightarrow 0$, we have
\begin{equation}
h(r)\rightarrow -\frac{8\sqrt{C}}{5\beta^{1/4}},~~~~h'(r)\rightarrow \frac{1}{\sqrt{\beta}}.
\label{13}
\end{equation}
In electrostatics $E=h'(r)$, and maximum field at the origin is $E_{max}=1/\sqrt{\beta}$ which is in agreement with \cite{Kruglov}. Thus, electric field has the finite value at the origin and there are no singularities.
The function $f(r)$ entering the spherically symmetric line element (6) can be found by the relation \cite{Capozziello}
\begin{equation}
f(r)=1+\frac{k_1}{r}+\frac{k_2}{r^2}+\frac{1}{r^2}\int dr\left[\int r^2R(r)dr\right],
\label{14}
\end{equation}
where $k_1,k_2$ are integration constants. To obtain the Ricci scalar we use the relation which follows from Einstein's equation (4):
\begin{equation}
R=-\kappa^2T,~~~~T=g^{\mu\nu}T_{\mu\nu}.
\label{15}
\end{equation}
From Eq. (2) we obtain the trace of the energy-momentum tensor
\begin{equation}
T=\frac{8\beta {\cal F}^2}{(1+2\beta {\cal F})^2}=\frac{2\beta (h'(r))^4}{[1-\beta (h'(r))^2]^2}.
\label{16}
\end{equation}
Then from Eqs. (15),(16) and (8) one finds the Ricci scalar
\begin{equation}
R=-\frac{2C\kappa^2\beta(h'(r))^3}{r^2}.
\label{17}
\end{equation}
Replacing $R$ from Eq. (17) into Eq. (14) we obtain, after integration, the function
\begin{equation}
f(r)=1+\frac{k_1}{r}+\frac{k_2}{r^2}+\frac{C^2\kappa^2}{30r^2}\left(5y^3-22y^2+32y\right),
\label{18}
\end{equation}
where $y$ is the solution to Eq. (10). The last terms in Eq. (18), containing the variable $y$, give
corrections to Reissner-Nordstr\"om solution.
%From Eq. (18) we obtain the cosmological constant generated by nonlinear electromagnetic field
%\begin{equation}
%\Lambda=\frac{\kappa^2}{2\beta}.
%\label{19}
%\end{equation}
%At $r\rightarrow 0$, we have from Eq. (18)
%\begin{equation}
%f(r)\rightarrow 1+ \frac{k_1}{r}+\frac{k'_2}{r^2},
%\label{20}
%\end{equation} where the cosmological constant $\Lambda$ is given by Eq. (19).
%where $k'$ is a new constant, and
At $r\rightarrow \infty$
\begin{equation}
f(r)\rightarrow 1+\frac{k_1}{r}+\frac{k_2}{r^2}+\frac{C^2\kappa^2}{2r^2}.
\label{19}
\end{equation}
and we can identify the charge of the black hole with the value $G^2Q^2=k_2+C^2\kappa^2/2$, where $G$ is the Newton constant. Asymptotically spacetime becomes flat. Thus, the metric function at $r\rightarrow \infty$ approaches
\begin{equation}
f(r)\rightarrow 1-\frac{2GM}{r}+\frac{G^2Q^2}{r^2},
\label{20}
\end{equation}
where $M$ is the mass of the black hole ($k_1=-2GM$). In the case of linear Maxwell's electrodynamics the parameter $\beta$ is zero, $\beta=0$. In this case from Eq. (9) we find that $x\rightarrow \infty$ and $y=1$. As a result, we come from Eq. (18) to the Reissner-Nordstr\"om solution.

\section{Conclusion}

We have considered a new model of nonlinear electromagnetic fields with a dimensional parameter $\beta$ in the Einstein gravity. The scale invariance is broken in this model and, as a result, the trace of the energy-momentum tensor is not zero. The static spherically symmetric solutions were obtained describing the charged black hole. The nonlinear electromagnetic field has a finite value $E_{max}=1/\sqrt{\beta}$ in the origin of the charged particle. The black hole solution obtained possesses the asymptotic Reissner-Nordstr\"om solution. At the same time in the limiting case of linear electrodynamics ($\beta \rightarrow 0$), we have the Reissner-Nordstr\"om solution. The electromagnetic fields can be coupled with gravity non-linearly \cite{Odintsov}.
It is also possible to generalize the study of current theory by introducing non-minimal coupling with the gravitation field. It is interesting to investigate the instabilities of RN black hole and the anti-evaporation effect in the model under consideration similar to the study of these effects in the Maxwell-$F(R)$ theory \cite{Odintsov}. We leave such investigations for the further consideration.

\end{document}